# Deeper understandings of the gauge theory for the first order inhomogeneous linear elasticity

Zhihai Xiang

*Department of Engineering Mechanics, Tsinghua University, Beijing 100084, China*

Email: xiangzhihai@tsinghua.edu.cn

**Abstract** Our previous study [1] has demonstrated that the gauge theory is a proper framework for characterizing the local temporal and spatial interactions in inhomogeneous elastic media. However, in that study temporal interactions were interpreted as the compensation for the loss of kinetic energy resulting from homogenization process, distinct from damping effects. In addition, that study did not account for the integration of temporal and spatial transformations, leading to the omission of some crucial information such as thermal stresses. In this paper, we address this oversight to establish generalized equations by employing a unified methodology that encompasses the integrated temporal-spatial transformations and the principle of minimum dissipation. Among many interesting new findings, we highlight that the newly derived equations are inherently consistent with the first and the second laws of thermodynamics, because this gauge theory naturally incorporates the fundamental mechanism that governs the partitioning between the dissipative and the non-dissipative energy.

# Nomenclature

| | |
|---|---|
| $A$ | Hamiltonian action |
| $b_\alpha$ | incompatibility vector |
| $^0B$ | initial configuration |
| $^\tau B$ | deformed configuration |
| $C$, $C_{ijkl}$ | elasticity tensor |
| $C^{-1}_{iikl}$ | components of compliance tensor |
| $\boldsymbol{C}^{\text{eff}}$ | elasticity operator in classical Willis equations |
| $\boldsymbol{e}$ | linear strain |
| $^0 e_{ij}$ | pre-strain |
| $\boldsymbol{e}_\alpha$ | basis of the global Cartesian time-space coordinate system |
| $E_\alpha$ | Euler-Lagrange expression |
| $\boldsymbol{f}, f_i$ | incremental volume density of external body force |
| $^0 f_i$ | volume density of external body force in $^0 B$ |
| $^0 f_i^E$, $^\tau f_i^E$ | volume density of effective body forces in $^0 B$ and $^\tau B$, respectively |
| $\boldsymbol{g}_\alpha$ | basis in the material manifold |
| $\bar{\boldsymbol{g}}_\gamma$ | basis in the material manifold projected onto $^0 B$ |
| $i$ | lower-case Roman indices represent spatial dimensions |
| $J_\beta$ | Noether current |
| $k_{\text{B}}$ | Boltzmann constant |
| $L$ | Lagrangian for homogeneous media |
| $L'$ | Lagrangian for inhomogeneous media |
| $\mathscr{L}$ | integration contour |
| $n_i$ | outward normal vector of $\partial V$ |
| $^\tau p_i$ | volume density of effective momentum |
| $R$ | incompatibility factor |
| $r_{\alpha\beta\mu}$ | incompatibility tensor |
| $R^\gamma_{\alpha\beta\mu}$ | Riemann curvature tensor |
| $S$ | surface |
| $\boldsymbol{S}^{\text{eff}}$ | velocity coupling operator in classical Willis equations |
| $\boldsymbol{S}^{\dagger\text{eff}}$ | adjoint operator of $\boldsymbol{S}^{\text{eff}}$ |
| $t$ | time |
| $T$ | incremental volume density of kinetic energy |
| $t_{\beta\gamma}$ | pseudo-tensor |
| $T_{\beta\gamma}$ | energy-momentum tensor |
| $\boldsymbol{u}, u_\alpha$ | temporal or spatial displacement of $\boldsymbol{X}$ |
| $\boldsymbol{u}', u'_\alpha$ | temporal or spatial displacement of $\boldsymbol{X}'$ |

| | | |
|---|---|---|
| $\bar{u}_\alpha$ | | distance from $X$ to $X'$ in $^0B$ |
| $^0_0 u_{i,j}$ | | gradient of the pre-displacement defined by $^0 e_{ij} = \left(^0_0 u_{i,j} + ^0_0 u_{j,i}\right)/2$ |
| $V$ | | spatial domain |
| $\partial V$ | | boundary of $V$ |
| $\bar{V}$ | | representative spatial domain of a material point |
| $W$ | | incremental volume density of strain energy |
| $^0W$ | | volume density of strain energy in $^0B$ |
| $^0x_\alpha$ | | temporal-spatial coordinates of $X$ in $^0B$ |
| $^0x'_\alpha$, $^\tau x'_\alpha$ | | temporal-spatial coordinate of $X'$ in $^0B$ and $^\tau B$, respectively |
| $X, X'$ | | two associated material points |
| $\alpha$ | | lower-case Greek indices represent temporal-spatial dimensions |
| $\alpha_{kl}$ | | thermal expansion tensor |
| $\Gamma^\gamma_{\alpha\beta}$ | | gauge potential |
| $\delta_{\alpha\beta}$ | | Kronecker delta symbol |
| $\varepsilon_{\alpha\beta\gamma}$ | | Levi-Civita symbol |
| $\theta$ | | temperature |
| $\Delta\theta$ | | incremental temperature |
| $\xi_i$ | | dependent parameter in global transformations |
| $\boldsymbol{\rho}$, $\rho_{ij}$ | | mass density tensor |
| $\boldsymbol{\rho}^{\text{eff}}$ | | density operator in classical Willis equations |
| $\boldsymbol{\sigma}, \sigma_{ij}$ | | incremental nominal stress |
| $^\tau\sigma_{ij}$ | | effective nominal stress in $^\tau B$ |
| $^0\boldsymbol{\sigma}$, $^0\sigma_{ij}$ | | pre-stress |
| $\Pi$ | | incremental volume densities of potential energy |
| $\Upsilon^\gamma_\alpha$ | | gauge field |
| $\Phi$ | | incremental volume density of the potential of external body forces |
| $\omega^\gamma_{\alpha\beta}$ | | torsion tensor |
| $\Omega$ | | temporal-spatial domain |

# 1 Introduction

Under the assumption of homogeneity [2], it is not necessary to consider the impact of local interactions in the initial configuration on the deformed configuration. Thereby, the theory of linear elasticity can be greatly simplified. However, local interactions are crucial for inhomogeneous materials.

The spatial gauge theory, introduced by Kondo [3], Bilby [4], and Kröner [5], was developed to characterize materials with micro-defects [6, 7]. This framework highlights the invariance of the Lagrangian under local Euclidean transformations. Thus, it extends *the principle of classical objectivity*, which only accounts for the invariance under global Euclidean transformations [8, 9]. Within this theory, local spatial interactions are depicted through pre-strains or pre-stresses caused by micro-defects, leading to additional terms in the generalized equations. As these concepts are inherently geometric, the theory is also known as geometric continuum mechanics [10-13]. Despite its limited popularity in engineering due to the high barrier of mastering advanced differential geometry, it has recently been utilized to model the evolutions of brittle [14] and ductile damage [15].

Besides the local spatial interaction mentioned above, the local temporal interaction was noticed by Willis over four decades ago [16]. He established the following equations for composites by using the ensemble average operation (denoted by $\langle \ \rangle$), together with his dynamic Green's function and the extension of Hashin–Shtrikman variational principle [17, 18]

$$\langle \boldsymbol{\sigma} \rangle = \boldsymbol{C}^{\text{eff}} * \langle \boldsymbol{e} \rangle + \boldsymbol{S}^{\text{eff}} \circ \langle \dot{\boldsymbol{u}} \rangle, \tag{1a}$$

$$\nabla \cdot \langle \boldsymbol{\sigma} \rangle + \boldsymbol{f} = \boldsymbol{S}^{\dagger\text{eff}} \circ \langle \dot{\boldsymbol{e}} \rangle + \boldsymbol{\rho}^{\text{eff}} \odot \langle \ddot{\boldsymbol{u}} \rangle, \tag{1b}$$

where $\boldsymbol{\sigma}$ is the nominal stress; $\boldsymbol{u}$ is the spatial displacement; $\boldsymbol{e} = (\boldsymbol{u}\nabla + \nabla\boldsymbol{u})/2$ is the linear strain; $\nabla$ is the spatial gradient operator; $\boldsymbol{f}$ is the incremental volume density of body force; the overhead dot denotes the time derivative; $\nabla\cdot$ is the spatial divergence operator; $\boldsymbol{C}^{\text{eff}}$ is the stiffness operator; $\boldsymbol{\rho}^{\text{eff}}$ is the density operator, which is generally in tensorial form due to the homogenization of constitutes inside an effective material point [19-21]; $*$, $\circ$ and $\odot$ are corresponding temporal-spatial convolutions; $\boldsymbol{S}^{\text{eff}} \circ \langle \dot{\boldsymbol{u}} \rangle$ is called the Willis coupling term; and $\boldsymbol{S}^{\dagger\text{eff}}$ is the adjoint operator of $\boldsymbol{S}^{\text{eff}}$. Unfortunately, similar to geometric continuum mechanics, this theory had not been taken seriously in engineering for a long time, probably due to their highly abstract formulas.

The situation changed after the rise of using the transformation method to design metamaterials [22]. With this method, the effective properties of metamaterials are obtained by comparing the corresponding terms in the original and transformed equations if they have the same form. Although this method had achieved great successes in transformation optics [23] and transformation acoustics [24], it failed in transformation elastics, because the classical linear elastodynamic equations change to Eqs. (1ab) under the local spatial transformation in frequency domain [25]. Since Eqs. (1ab) play non-fungible roles to design elastic metamaterials, they had aroused much interest and are customarily called the Willis equations.

It is important to note that the Willis equations in the time domain are not unique [26, 27]. Eq. (1) features a velocity-coupled term with a real-valued temporal-spatial convolution operator $\boldsymbol{S}^{\text{eff}}$. Nonetheless, $\boldsymbol{S}^{\text{eff}}$ could alternatively be a complex-valued algebraic operator [28-31], given that dissipation is not taken into account. Additionally, the acceleration-coupled algebraic operator is a viable alternative with experimental demonstrations [32-34]. Furthermore, the displacement-coupled pre-stress gradient can be derived directly by applying local spatial transformations to the classical elastodynamic equations [35-37]

$$\boldsymbol{\sigma} = \boldsymbol{C}:\boldsymbol{e} + \left(^0\boldsymbol{\sigma}\nabla\right)\cdot\boldsymbol{u}, \tag{2a}$$

$$\nabla\cdot\boldsymbol{\sigma} + \boldsymbol{f} = \left(^0\boldsymbol{\sigma}\nabla\right)^{\text{T}}:\boldsymbol{e} + \left[\left(\nabla\cdot{}^0\boldsymbol{\sigma}\right)\nabla\right]\cdot\boldsymbol{u} + \boldsymbol{\rho}\cdot\ddot{\boldsymbol{u}}, \tag{2b}$$

where $\boldsymbol{C}$ is the elasticity tensor; $^0\boldsymbol{\sigma}$ is the pre-stress; $\boldsymbol{\rho}$ is the mass density tensor. Eq. (2) has been experimentally validated [38, 39] and subsequently applied to the buckling analysis of thin-walled shells [40, 41].

In the previous study [1], we proved that the non-uniqueness of the Willis equations can be understood through local gauge transformations. The classical Willis equations given in Eq. (1) are the result of local temporal gauge transformation. The Willis-form equations given in Eq. (2) are the result of local spatial gauge transformation, similar to geometric continuum mechanics. However, the velocity coupled term $\boldsymbol{S}^{\text{eff}}\circ\langle\dot{\boldsymbol{u}}\rangle$ in Eq. (1) was simply regarded as the lost kinetic energy in the homogenization process. This did not give a satisfactory answer to the energy conservation of the classical Willis equations [42, 43]. In addition, the integrated temporal-spatial transformation was not discussed in that work, so that the gauge potentials were not complete.

Upon discussing the integrated temporal-spatial transformation, we have to clarify the definitions of time and space, which have plagued people for centuries. According to the review in a short book of Rovelli [44], Aristotle believed that time and space are relative concepts: time is only a way of measuring how things change; the place of a thing is what surrounds that thing. But in Newtonian mechanics, time and space are absolute entities, which are independent of things and uniform. Einstein synthesized the concepts of Aristotle and Newton. In the general theory of relativity, space and time correlate with gravity; they are coupled together and neither absolute nor uniform. Since time is not uniform, the clocks in the global positioning system must carefully account for gravitational frequency shifts.

Besides traditional time-space concepts, Rovelli also puts forward the audacious idea that time is discrete accounting for quantum effects [44]. This idea aligns with the thoughts of Nobel laureate

Lee, who had previously theorized that discrete time could represent a fundamental concept in physics [45, 46]. Simultaneously, there have been significant developments in the realm of discrete variational mechanics [47]. It is also noteworthy that our subjective experience of time [48] closely mirrors Aristotle's assertion that time is the measure of change. Anyway, the discussion on the essence of space-time is still inconclusive. However, all reasonings in this paper are grounded in macroscopic physical perspectives on time and space, excluding quantum, relativistic, and psychological factors. Within this context, space is perceived as the relative positioning of matter, and time signifies the change of space. Time and space are intertwined, with space having three dimensions and velocities considerably lower than the speed of light.

In continuum mechanics, the Aristotle's concept of "thing" is abstracted as a material point, which has stable statistically averaged macro properties contributed by group of micro particles (atoms or molecules) [49, 50]. In this classical model, the properties of a material point come from a representative domain, which can overlap with the representative domains of other material points. The overlapped part contains electro-magnetic interaction forces, that bind material points together to form a continuum possessing cohesive energy [51]. As will be elaborated in the subsequent text, the gauge field is intimately connected with these interactions. This classical model that accounts for the interaction of material points within the gauge field can also be envisioned as the interactions among micro-finite elements [52].

The world is a collection of events or change [44]. For a continuum, an event is equivalent to a state. To facilitate calculation, the I-th state of a material point is marked by coordinate $^{I}x_\alpha$ ($\alpha = t$, 1, 2, 3) in a global Cartesian time-space coordinate system with basis $e_\alpha$. Consequently, the I-th state of a continuum is represented by a set of such temporal-spatial coordinates, and is denoted by configuration $^{I}B$. In this paper, lower-case Greek indices, such as $\alpha$, represent both temporal dimension $t$ and spatial dimensions 1, 2 and 3; and lower-case Roman indices, such as $i$, only represent spatial dimensions. Since time is the measurement of change, $t$ can be regarded as the marker of states. For simplicity, we do not assign a velocity to the temporal dimension to maintain the same unit as spatial dimensions, because it does not impact the accuracy of the outcome.

During the change from an initial configuration $^{0}B$ to a deformed configuration $^{\tau}B$, the magnitude and direction of the movement of a material point are measured by temporal-spatial displacement

$$u_\alpha = {}^{\tau}x_\alpha - {}^{0}x_\alpha ; \qquad (3)$$

and the rate of the movement is measured by temporal-spatial velocity

$$^{0}_{0}x_{\alpha,\beta} = \lim_{\tau \to 0} \frac{{}^{\tau}x_\alpha - {}^{0}x_\alpha}{{}^{\tau}x_\beta - {}^{0}x_\beta} = \frac{\partial {}^{0}x_\alpha}{\partial {}^{0}x_\beta}. \qquad (4)$$

In this paper, the comma in the right subscript signifies the application of the partial derivative operator; the left subscript denotes the specific state of the coordinate undergoing differentiation. With this definition, it is easy to find $_0^0x_{i,j} = \delta_{ij}$ ($\delta_{ij}$ is the Kronecker symbol). $_0^0x_{i,t}$ is the classical velocity, indicating the change in spatial position of a material point per unit time. The reciprocal of $_0^0x_{i,t}$ is slowness $_0^0x_{t,i}$, indicating the time experienced through unit spatial distance. If $_0^0x_{i,t}$ is spatially nonuniform, $_0^0x_{t,i}$ is also spatially nonuniform, implying the spatial non-uniformity of time. Generally $_0^0x_{t,t} \neq 1$, indicating the temporal non-uniformity of time. The temporal-spatial non-uniformity of time is not solely a consequence of gravity [44]; it extends to classical mechanics as well [45, 46], underscoring the fundamental comprehension of time as a dynamic variable. Similarly, we can define temporal-spatial acceleration and jerk, etc., which are not further elaborated here.

Based on the above understanding of time and space, the rest of this paper explores the theory of the first order linear elasticity for inhomogeneous media by using the integrated temporal-spatial local transformations. Section 2 examines temporal-spatial gauge fields through the implementation of the minimal replacement technique [6, 7, 53]. This section not only addresses the mathematical aspects of the gauge field but also underscores its physical implications. Section 3 formulates a generalized version of the linear elastodynamic equations by incorporating temporal-spatial gauge fields based on *the Principle of Least Action* (PLA) [54], with the gauge field being further determined by *the Principle of Minimum Dissipation* (PMD) [55]. Section 4 explores the relationship between the generalized equations and the theory of thermodynamics. Finally, Section 5 presents conclusions and future perspectives.

## 2 Gauge Fields

### 2.1 Minimal Replacement Method

In accordance with the PLA [54], the Hamiltonian action from the initial configuration $^0B$ to a deformed configuration should achieve an extreme value. The first order version of this action can be generally denoted as

$$A = \int_{\Omega} L\left(^0x_\alpha, u_\alpha, {_0}u_{\alpha,\beta}\right) \mathrm{d}\Omega, \tag{5}$$

where $\Omega$ is a temporal-spatial domain represented in the global Cartesian time-space coordinate system; $^0x_\alpha$ is the independent variable; and dependent variables are chosen as the temporal-spatial displacement $u_\alpha$ and its first derivative $_0u_{\alpha,\beta}$, because these quantities can be directly measured. For a conservative mechanical system without dissipation, the Lagrangian of this deformation process is defined as

$$L\left({}^{0}x_{\alpha},u_{\alpha},{}_{0}u_{\alpha,\beta}\right)=T-\Pi. \tag{6}$$

where $T$ and $\Pi$ are incremental volume densities of kinetic energy and potential energy, respectively. Since the mass density is a symmetric tensor, $T$ is defined as

$$T=\frac{1}{2}{}^{\tau}\rho_{ij}\,{}^{\tau}_{\tau}x_{i,t}\,{}^{\tau}_{\tau}x_{j,t}-\frac{1}{2}{}^{0}\rho_{ij}\,{}^{0}_{0}x_{i,t}\,{}^{0}_{0}x_{j,t}, \tag{7a}$$

where we use Einstein's notation of index summation. In linear elasticity, we consider only infinitesimal perturbations and represent all quantities in the initial configuration ${}^0B$. Therefore, ${}^{\tau}\rho_{ij}=\det\left({}^{\tau}_{0}x_{k,l}\right){}^{0}\rho_{ij}\approx\left(1+{}_{0}u_{k,k}\right){}^{0}\rho_{ij}\approx{}^{0}\rho_{ij}$ and ${}^{\tau}_{\tau}x_{i,t}={}_{\tau}\left({}^{0}x_{i}+u_{i}\right)_{,t}\approx{}^{0}_{0}x_{i,t}+{}_{0}u_{i,t}$. Hence, Eq. (7a) can be rewritten as

$$T\approx{}^{0}\rho_{ij}\,{}^{0}_{0}x_{i,t}\,{}_{0}u_{j,t}+\frac{1}{2}{}^{0}\rho_{ij}\,{}_{0}u_{i,t}\,{}_{0}u_{j,t}. \tag{7b}$$

Considering the contributions of the pre-stress ${}^{0}\sigma_{ij}$ and the volume density of external body force ${}^{0}f_{i}$ in ${}^0B$, the definition of $\Pi$ for a homogeneous medium is

$$\Pi=\underbrace{{}^{0}\sigma_{ij}\,{}_{0}u_{i,j}+\frac{1}{2}C_{ijkl}\,{}_{0}u_{i,j}\,{}_{0}u_{k,l}}_{W}-\underbrace{\left({}^{0}f_{i}+\frac{1}{2}{}^{0}_{0}f_{i,j}u_{j}+f_{i}\right)u_{i}}_{\Phi}, \tag{8}$$

where $W$ and $\Phi$ denote incremental volume densities of strain energy and the potential of external body force, respectively. The elasticity tensor is widely recognized to possess major symmetry, which is necessitated by the conservation of energy [2]. Since we do not consider local body torque (e.g., electromagnetic couples [49] or polar materials [56]), stress tensors are symmetric. Consequently, the elasticity tensor has both major and minor symmetries, i.e., $C_{ijkl}=C_{klij}=C_{jikl}=C_{ijlk}$.

Because the laws of nature are independent of the choice of coordinate systems, the action in Eq. (5) should be the same when the integration is conducted in another arbitrary coordinate system marked by a prime

$$A=\int_{\Omega}L\left({}^{0}x_{\alpha},u_{\alpha},{}_{0}u_{\alpha,\beta}\right)\mathrm{d}\Omega=\int_{\Omega'}L\left({}^{\tau}x',u'_{\alpha},{}_{\tau}u'_{\alpha,\beta'}\right)\mathrm{d}\Omega'. \tag{9}$$

The change of coordinate system means the change of view point. Therefore, Eq. (9) implies that the action $A$ is invariant after the view point moves from a material point $X$ in ${}^0B$ to an infinitely close (in the sense of time and space) material point $X'$ in ${}^{\tau}B$, i.e., through the following coordinate transformation

$$X\left({}^{0}x_{\alpha}\right)\;\;\rightarrow\;\;X'\left({}^{\tau}x'_{\alpha}\right). \tag{10}$$

Upon the change of view point, the observed displacement also changes, so that we have the following gauge transformation

$$\boldsymbol{u}(^{0}x_{\alpha}) \rightarrow \boldsymbol{u}'(^{\tau}x'_{\alpha}). \tag{11}$$

With these notations, $_{\tau}u'_{\alpha,\beta'}$ in Eq. (9) signifies $\partial u'_{\alpha}/\partial\,^{\tau}x'_{\beta}$.

According to the coordinate transformation given in Eq. (10), the differential vector of material point $X'$ can be written as $\mathrm{d}^{\tau}\boldsymbol{r} = \mathrm{d}^{\tau}x'_{\gamma}\boldsymbol{e}_{\gamma} = \mathrm{d}^{0}x_{\alpha}\boldsymbol{g}_{\alpha}$, where $\boldsymbol{g}_{\alpha}$ is the basis in the material manifold

$$\boldsymbol{g}_{\alpha} = {}^{\tau}_{0}x'_{\gamma,\alpha}\boldsymbol{e}_{\gamma}. \tag{12}$$

In addition, the gauge transformation given in Eq. (11) specifies $u_{\gamma}\boldsymbol{e}_{\gamma} = u'_{\alpha}\boldsymbol{g}_{\alpha}$. Referring to Eq. (12), we have the following mapping

$$u'_{\alpha} = {}^{0}_{\tau}x_{\gamma,\alpha'}u_{\gamma}. \tag{13}$$

Eq. (13) represents the Lodge transformation [57]. As noted in [58], this transformation was independently employed in [25] for a distinct purpose.

The interactions in homogeneous media are homogeneous, so that we can 'see' all properties by using only global transformations. In this case, ${}^{0}_{\tau}x_{\gamma,\alpha'}$ is independent of coordinates, and Eq. (9) can be easily satisfied because $L$ is form-invariant, i.e., $L\left({}^{\tau}x',u'_{\alpha},{}_{\tau}u'_{\alpha,\beta'}\right)=L\left({}^{\tau}x',u'_{\alpha},{}^{0}_{\tau}x_{\gamma,\alpha'}\,_{\tau}u_{\gamma,\beta'}\right)$. This is *the principle of classical objectivity* [8, 9].

The interactions in inhomogeneous media are inhomogeneous, so that we need local transformations to find all details. In this case, ${}^{0}_{\tau}x_{\gamma,\alpha'}$ varies with coordinates, and $L$ does change its form: $L\left({}^{\tau}x',u'_{\alpha},{}_{\tau}u'_{\alpha,\beta'}\right)=L\left({}^{\tau}x',u'_{\alpha},{}^{0}_{\tau}x_{\gamma,\alpha'\beta'}u_{\gamma}+{}^{0}_{\tau}x_{\gamma,\alpha'}\,_{\tau}u_{\gamma,\beta'}\right)$. According to *the principle of generalized objectivity* [7], if Eq. (9) is still valid we have to use the minimal replacement method [53] to construct a form-invariant $L'$:

$$L'\left({}^{\tau}x',u'_{\alpha},{}_{\tau}u'_{\alpha,\beta'}\right)=L\left({}^{\tau}x',u'_{\alpha},{}_{\tau}u'_{\alpha;\beta'}\right)=L\left({}^{\tau}x',u'_{\alpha},{}^{0}_{\tau}x_{\gamma,\alpha'}\,_{\tau}u_{\gamma;\beta'}\right), \tag{14}$$

where the semicolon denotes the covariant derivative that satisfies

$$_{\tau}u'_{\alpha;\beta'} = {}_{\tau}\left({}^{0}_{\tau}x_{\gamma,\alpha'}u_{\gamma}\right)_{;\beta'} = {}^{0}_{\tau}x_{\gamma,\alpha'}\,_{\tau}u_{\gamma;\beta'} = {}^{0}_{\tau}x_{\gamma,\alpha'}\,_{0}u_{\gamma;\mu}\,^{\tau}_{0}x_{\mu,\beta'}. \tag{15}$$

The general form of covariant derivative of the displacement is

$${}_{0}u_{\alpha;\beta} = {}_{0}u_{\alpha,\beta} + G_{\alpha\beta}. \tag{16}$$

$G_{\alpha\beta}$ acts as the connection between a material point and its neighboring material point during the change of view point described in Eq. (10). The physical significance of this minimal replacement operation is that energy should remain unaffected by changes in view point, such that the partial derivative cannot independently contribute to the evaluation of the Lagrangian [7]. $G_{\alpha\beta}$ is not a tensor, but satisfies the coordinate transformation relationship given in Appendix A, which ensures the validity of Eq. (15).

The specific formulation of $_0u_{\alpha;\beta}$ is determined by the difference between $\boldsymbol{u}'$ and $\boldsymbol{u}$. Given that $\boldsymbol{u}'$ and $\boldsymbol{u}$ belong to different configurations, it is necessary to project $\boldsymbol{u}'$ onto the initial configuration $^0B$. This projection allows for a direct comparison between $\boldsymbol{u}'$ and $\boldsymbol{u}$ [59]. The projected $\boldsymbol{u}'$ is denoted as $\bar{\boldsymbol{u}}' = (u_\gamma + {_0u_{\gamma,\beta}}\mathrm{d}^0 x_\beta)\bar{\boldsymbol{g}}_\gamma$. According to Eq. (12), $\bar{\boldsymbol{g}}_\gamma$ is simply

$$\bar{\boldsymbol{g}}_\gamma = {^0x'_{\alpha,\gamma}}\boldsymbol{e}_\alpha, \tag{17}$$

where $^0x'_\alpha$ is the coordinate of material point $\boldsymbol{X}'$ in configuration $^0B$. The distance from $\boldsymbol{X}$ to $\boldsymbol{X}'$ in configuration $^0B$ is

$$\bar{u}_\alpha({^0x'_\alpha}) = {^0x'_\alpha} - {^0x_\alpha}. \tag{18}$$

Substituting Eq. (18) into Eq. (17), we obtain

$$\bar{\boldsymbol{g}}_\gamma = (\delta_{\alpha\gamma} + {_0\bar{u}_{\alpha,\gamma}})\boldsymbol{e}_\alpha. \tag{19}$$

Upon observing $\bar{u}_\alpha({^0x'_\alpha}) \approx \bar{u}_\alpha({^0x_\alpha}) + {_0\bar{u}_{\alpha,\beta}}({^0x_\alpha})\mathrm{d}^0 x_\beta = {_0\bar{u}_{\alpha,\beta}}({^0x_\alpha})\mathrm{d}^0 x_\beta$, it follows that

$$_0\bar{u}_{\alpha,\gamma} \approx {_0\bar{u}_{\alpha,\beta\gamma}}\mathrm{d}^0 x_\beta. \tag{20}$$

With above discussions, we can calculate the difference between $\boldsymbol{u}'$ and $\boldsymbol{u}$

$$\begin{aligned}\bar{\boldsymbol{u}}' - \boldsymbol{u} &= (u_\gamma + {_0u_{\gamma,\beta}}\mathrm{d}^0 x_\beta)(\delta_{\alpha\gamma} + {_0\bar{u}_{\alpha,\gamma}})\boldsymbol{e}_\alpha - u_\alpha \boldsymbol{e}_\alpha \\ &= \left[{_0\bar{u}_{\alpha,\gamma}}u_\gamma + ({_0u_{\alpha,\beta}} + {_0\bar{u}_{\alpha,\gamma}}{_0u_{\gamma,\beta}})\mathrm{d}^0 x_\beta\right]\boldsymbol{e}_\alpha \\ &\approx ({_0u_{\alpha,\beta}} + {_0\bar{u}_{\alpha,\beta\gamma}}u_\gamma + {_0\bar{u}_{\alpha,\gamma}}{_0u_{\gamma,\beta}})\mathrm{d}^0 x_\beta \boldsymbol{e}_\alpha\end{aligned} \tag{21}$$

Comparing Eq. (21) with Eq. (16), we obtain

$$_0u_{\alpha;\beta} = {_0u_{\alpha,\beta}} + {_0\bar{u}_{\alpha,\beta\gamma}}u_\gamma + {_0\bar{u}_{\alpha,\gamma}}{_0u_{\gamma,\beta}} = {_0u_{\alpha,\beta}} + \Gamma^\gamma_{\alpha\beta}u_\gamma + \Upsilon^\gamma_\alpha {_0u_{\gamma,\beta}}, \tag{22}$$

$$G_{\alpha\beta} = \Gamma^\gamma_{\alpha\beta}u_\gamma + \Upsilon^\gamma_\alpha {_0u_{\gamma,\beta}} = {_0({_0\bar{u}_{\alpha,\gamma}}u_\gamma)_{,\beta}}. \tag{23}$$

$\Upsilon^\gamma_\alpha = {_0\bar{u}_{\alpha,\gamma}}$ is the gauge field. $\Gamma^\gamma_{\alpha\beta} = {_0\bar{u}_{\alpha,\beta\gamma}}$ is called the gauge potential. It is interesting to note that unlike classical differential geometry [59], the covariant derivative in Eq. (22) includes a generalized connection that encompasses not only the pure rigid body transformation (or parallel transport in terms of differential geometry) represented by $\Gamma^\gamma_{\alpha\beta}u_\gamma$ [7], but also the stretch and shear represented by $\Upsilon^\gamma_\alpha {_0u_{\gamma,\beta}}$.

Given that the material point $\boldsymbol{X}'$ is chosen arbitrarily, the gauge field $_0\bar{u}_{\alpha,\gamma}$ is not unique. Consequently, there are gauge freedoms, a significant characteristic of gauge theories [60, 61]. The gauge field can be fixed under a specific gauge condition, which endows it with clear physical implications. This concept will be explored in more detail in Section 3.2.

It should be noted that $\Gamma^\gamma_{\alpha\beta} \neq \Gamma^\gamma_{\beta\alpha}$, which implies the existence of Cartan's torsion [7, 10-12,

62]

$$\omega^{\gamma}_{\alpha\beta} = \Gamma^{\gamma}_{\alpha\beta} - \Gamma^{\gamma}_{\beta\alpha} = {_0}\left({_0\bar{u}_{\alpha,\beta}} - {_0\bar{u}_{\beta,\alpha}}\right)_{,\gamma}. \tag{24}$$

## 2.2 Temporal-Spatial Incompatibility

Based on Eq. (22), we can calculate the following temporal-spatial incompatibility vector through contour integration and the Stokes theorem:

$$b_\alpha = \oint_{\mathscr{L}} {_0 u_{\alpha;\beta}} d^0 x_\beta = -\int_S \varepsilon_{\beta\mu\nu}\, {_0}\left({_0 u_{\alpha;\beta}}\right)_{;\mu} dS = -\frac{1}{2}\int_S \varepsilon_{\beta\mu\nu}\left[{_0}\left({_0 u_{\alpha;\beta}}\right)_{;\mu} - {_0}\left({_0 u_{\alpha;\mu}}\right)_{;\beta}\right] dS, \tag{25}$$

where $\varepsilon_{\beta\mu\nu}$ is the Levi-Civita symbol; and $S$ is a surface bounded by contour $\mathscr{L}$.

According to Eqs. (22) and (23), the second covariant derivative of displacement is

$$\begin{aligned}
{_0}\left({_0 u_{\alpha;\beta}}\right)_{;\mu} &= {_0}\left({_0 u_{\alpha,\beta}} + {_0 G_{\alpha\beta}}\right)_{;\mu} \\
&= {_0}\left({_0 u_{\alpha,\beta}} + {_0 G_{\alpha\beta}}\right)_{,\mu} + \Gamma^{\gamma}_{\alpha\mu}\left({_0 u_{\gamma,\beta}} + G_{\gamma\beta}\right) + \Gamma^{\gamma}_{\beta\mu}\left({_0 u_{\alpha,\gamma}} + G_{\alpha\gamma}\right) \\
&\quad + \Upsilon^{\gamma}_{\alpha}\left({_0 u_{\gamma,\beta\mu}} + {_0 G_{\gamma\beta,\mu}}\right) + \Upsilon^{\gamma}_{\beta}\left({_0 u_{\alpha,\gamma\mu}} + {_0 G_{\alpha\gamma,\mu}}\right)
\end{aligned} \tag{26}$$

Therefore, the incompatibility tensor in Eq. (25) can be calculated as

$$\begin{aligned}
r_{\alpha\beta\mu} &= {_0}\left({_0 u_{\alpha;\beta}}\right)_{,\mu} - {_0}\left({_0 u_{\alpha;\mu}}\right)_{,\beta} \\
&= \left({_0\Gamma^{\gamma}_{\alpha\beta,\mu}} - {_0\Gamma^{\gamma}_{\alpha\mu,\beta}}\right) u_\gamma + \Gamma^{\gamma}_{\alpha\mu} G_{\gamma\beta} - \Gamma^{\gamma}_{\alpha\beta} G_{\gamma\mu} + \omega^{\gamma}_{\beta\mu}\, {_0 u_{\alpha;\gamma}} \\
&\quad + {_0\Upsilon^{\gamma}_{\alpha,\mu}}\, {_0 u_{\gamma,\beta}} - {_0\Upsilon^{\gamma}_{\alpha,\beta}}\, {_0 u_{\gamma,\mu}} + \Upsilon^{\gamma}_{\beta}\, {_0}\left({_0 u_{\alpha;\gamma}}\right)_{,\mu} - \Upsilon^{\gamma}_{\mu}\, {_0}\left({_0 u_{\alpha;\gamma}}\right)_{,\beta}
\end{aligned} \tag{27}$$

Under pure translation, $\Upsilon^{\gamma}_{\alpha}$ is neglected and $G_{\gamma\mu} = \Gamma^{\nu}_{\gamma\mu} u_\nu$. Thus, Eq. (27) becomes

$$r_{\alpha\beta\mu} = R^{\gamma}_{\alpha\beta\mu} u_\gamma + \omega^{\gamma}_{\beta\mu}\, {_0 u_{\alpha;\gamma}}, \tag{28}$$

where $R^{\gamma}_{\alpha\beta\mu} = {_0\Gamma^{\gamma}_{\alpha\beta,\mu}} - {_0\Gamma^{\gamma}_{\alpha\mu,\beta}} + \Gamma^{\nu}_{\alpha\mu}\Gamma^{\gamma}_{\nu\beta} - \Gamma^{\nu}_{\alpha\beta}\Gamma^{\gamma}_{\nu\mu}$. Although $R^{\gamma}_{\alpha\beta\mu}$ has the similar form of the Riemann curvature tensor [63], they have substantial differences. Because $\Gamma^{\gamma}_{\alpha\beta} = {_0\bar{u}_{\alpha,\beta\gamma}}$, it is easy to find that ${_0\Gamma^{\gamma}_{\alpha\beta,\mu}} = {_0\Gamma^{\gamma}_{\alpha\mu,\beta}}$ and $\omega^{\gamma}_{\beta\mu} \neq 0$. On the contrary in the classical Riemannian geometry, ${_0\Gamma^{\gamma}_{\alpha\beta,\mu}} \neq {_0\Gamma^{\gamma}_{\alpha\mu,\beta}}$ and $\omega^{\gamma}_{\beta\mu} = 0$.

When focusing solely on spatial dimensions, the incompatibility vector $b_i$ correlates with the incompatibility induced by micro-defects, such as Burgers and Frank vectors [6, 10-12, 63]. Section 3.2 will elucidate that the pure spatial gauge ${_0\bar{u}_{i,j}}$ is associated with the pre-strain, providing a natural representation of spatial incompatibilities. Moreover, $b_i$ also contains temporal incompatibility, which is associated to the pure temporal gauge ${_0\bar{u}_{i,t}}$. The relative velocity among material points is certainly associated with dissipation, as indicated in [64]. Comprehensive discussions on the relationship between the gauge theory and dissipations in time and space are scheduled for Section 4. The general form $b_\alpha$ encompasses even more complex incompatibilities related to the coupling effects of time and space.

## 2.3 Time-Space Conversion

Temporal and spatial components of the covariant derivative given in Eq. (22) can be detailed as

$$_0u_{i;t} = {}_0u_{i,t} + {}_0\left({}_0\bar{u}_{i,k}u_k\right)_{,t} + {}_0\left({}_0\bar{u}_{i,t}\mathrm{d}t\right)_{,t}, \tag{29a}$$

$$_0u_{i;j} = {}_0u_{i,j} + {}_0\left({}_0\bar{u}_{i,k}u_k\right)_{,j} + {}_0\left({}_0\bar{u}_{i,t}\mathrm{d}t\right)_{,j}, \tag{29b}$$

$$_0u_{t;t} = {}_0u_{t,t} + {}_0\left({}_0\bar{u}_{t,k}u_k\right)_{,t} + {}_0\left({}_0\bar{u}_{t,t}\mathrm{d}t\right)_{,t}, \tag{29c}$$

$$_0u_{t;i} = {}_0u_{t,i} + {}_0\left({}_0\bar{u}_{t,k}u_k\right)_{,i} + {}_0\left({}_0\bar{u}_{t,t}\mathrm{d}t\right)_{,i}, \tag{29d}$$

where $\mathrm{d}t$ denotes $u_t$, since the change of configuration is an infinitesimal perturbation. ${}_0\bar{u}_{i,t}$ and ${}_0\bar{u}_{i,j}$ are temporal and spatial gauges, respectively. ${}_0\bar{u}_{i,tt}$ results from pure temporal inhomogeneity, ${}_0\bar{u}_{i,jk}$ results from pure spatial inhomogeneity, and ${}_0\bar{u}_{i,tk} = {}_0\bar{u}_{i,kt}$ represents the temporal-spatial coupling effect.

Eqs. (29a) and (29b) are temporal and spatial covariant derivatives of spatial displacement, corresponding to velocity and strain, respectively. As discussed in the Introduction, time is nonuniform both in temporal and spatial dimensions [44-46]. Consequently, despite the unfamiliar appearances, Eqs. (29c) and (29d) are still comprehensible.

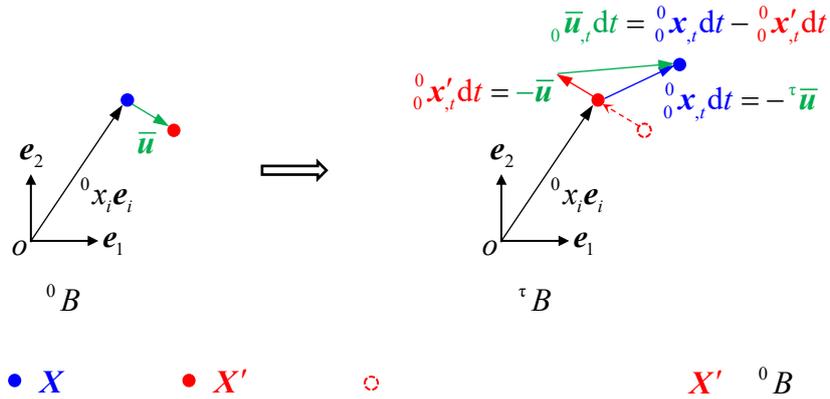

**Fig. 1** The positions of $X$ and $X'$ in different configurations are depicted in a simplified manner within a two-dimensional space.

However, the use of $\mathrm{d}t$ in Eq. (29) is not a conventional expression. Hence, it is better to convert ${}_0\bar{u}_{\alpha,t}\mathrm{d}t$ to a more comprehensible form. For this purpose, it is imperative to examine the change of $\bar{u}_\alpha$ at the spatial position ${}^0x_i\boldsymbol{e}_i$, which was initially occupied by the material point $X$ in ${}^0B$, because all variables in linear elasticity must be expressed in the initial configuration. As Fig. 1 illustrates, since the material point $X'$ associated with $X$ can be freely designated, we select the $X'$ that moves to the spatial position ${}^0x_i e_i$ in the deformed configuration ${}^\tau B$. Consequently, during the perturbation from ${}^0B$ to ${}^\tau B$, the displacement of $X'$ is $u'_\alpha = {}^0_0 x'_{\alpha,t} dt = -\bar{u}_\alpha$; whereas the displacement of $X$ is $u_\alpha = {}^0_0 x_{\alpha,t} dt = -{}^\tau\bar{u}_\alpha$, in which ${}^\tau\bar{u}_\alpha$ is the distance from $X$ to $X'$ in ${}^\tau B$. Because the material point at the spatial position ${}^0x_i e_i$ has changed from $X$ in ${}^0B$ to $X'$ in ${}^\tau B$, we have

$$_0\bar{u}_{\alpha,t} dt = -\left({}^\tau\bar{u}_\alpha - \bar{u}_\alpha\right) = u_\alpha - u'_\alpha \approx -_0 u_{\alpha,k} d\,{}^0x_k = -_0 u_{\alpha,k} \bar{u}_k. \tag{30}$$

With the time-space conversion relationship given in Eq. (30), Eqs. (29ab) can be rewritten as:

$$_0 u_{i;t} = {}_0 u_{i,t} + {}_0 \bar{u}_{i,tk} u_k + {}_0 \bar{u}_{i,k}\, {}_0 u_{k,t} - {}_0 \bar{u}_{k,t}\, {}_0 u_{i,k} - \bar{u}_k\, {}_0 u_{i,kt}, \tag{31a}$$

$$_0 u_{i;j} = {}_0 u_{i,j} + {}_0 \bar{u}_{i,jk} u_k + {}_0 \bar{u}_{i,k}\, {}_0 u_{k,j} - {}_0 \bar{u}_{k,j}\, {}_0 u_{i,k} - \bar{u}_k\, {}_0 u_{i,kj}. \tag{31b}$$

It is noteworthy that $_0 u_{i;t}$ contains $_0 u_{i,kt}$, which correlates with the stain rate, reflecting the time scale effect [65]; meanwhile, $_0 u_{i;j}$ encompasses $_0 u_{i,kj}$, which is related to the stain gradient, reflecting the length scale effect [66]. However, Eqs. (31a) and (31b) indicate that both time scale and length scale effects are also closely linked to the gauge field $_0 \bar{u}_{i,\alpha}$ and gauge potential $_0 \bar{u}_{i,\alpha k}$. Thus, it seems that using a single time scale or a single length scale to describe the temporal or spatial multi-scale problems might not be adequate.

If we adhere to the first order theory, we disregard $_0 u_{i,kt}$ and $_0 u_{i,kj}$. Furthermore, in linear elasticity, $_0 \bar{u}_{i,k} \ll 1$. Thus, Eq. (31) can be simplified as:

$$_0 u_{i;t} = {}_0 u_{i,t} + {}_0 \bar{u}_{i,tk} u_k - {}_0 \bar{u}_{k,t}\, {}_0 u_{i,k}, \tag{32a}$$

$$_0 u_{i;j} = {}_0 u_{i,j} + {}_0 \bar{u}_{i,jk} u_k. \tag{32b}$$

Comparing Eq. (32) with Eq. (27) from our previous paper [1], the only distinction lies in the extra term $_0 \bar{u}_{i,tk} u_k$ present in Eq. (32a), signifying the temporal-spatial coupling effect.

## 3 Generalized Elastodynamic Equations

### 3.1 Euler-Lagrange Equation

By extremizing the Hamiltonian action presented in Eq. (9) and taking into account the boundary condition where field variables vanish, we derive the Euler-Lagrange equation as detailed in reference [54]:

$$E_\alpha = \frac{\partial L'}{\partial u_\alpha} - \left(\frac{\partial L'}{\partial\, _0 u_{\alpha,\beta}}\right)_{,\beta} = 0, \tag{33}$$

where $E_\alpha$ can be called as Euler-Lagrange expression. The temporal and spatial decompositions of Eq. (33) are

$$\frac{\partial L'}{\partial u_t} - \left(\frac{\partial L'}{\partial \, _0 u_{t,\beta}}\right)_{,\beta} = \frac{\partial L'}{\partial u_t} - \left(\frac{\partial L'}{\partial \, _0 u_{t,j}}\right)_{,j} - \left(\frac{\partial L'}{\partial \, _0 u_{t,t}}\right)_{,t} = 0, \qquad (34)$$

$$\frac{\partial L'}{\partial u_i} - \left(\frac{\partial L'}{\partial \, _0 u_{i,\beta}}\right)_{,\beta} = \frac{\partial L'}{\partial u_i} - \left(\frac{\partial L'}{\partial \, _0 u_{i,j}}\right)_{,j} - \left(\frac{\partial L'}{\partial \, _0 u_{i,t}}\right)_{,t} = 0. \qquad (35)$$

Eq. (34) is the generalized condition of energy evolution [45, 46]. In Newtonian mechanics, time is considered to be uniform and gauge fields are not taken into account. Consequently, Eq. (34) simplifies to $_0 L_{,t} = 0$, which is the classical condition for energy conservation (see Appendix B.2). When focusing solely on the nonuniformity of time and disregarding gauge fields, Eq. (34) serves as the foundation for modern numerical integration techniques that employ adaptive time stepping [67].

Disregarding the nonuniformity of time, we can concentrate on Eq. (35), which represents the generalized condition of momentum evolution. The corresponding equation of motion is

$$_0^\tau \sigma_{ij,j} + {}^\tau f_i^E = {}^\tau \dot{p}_i, \qquad (36)$$

where ${}^\tau \sigma_{ij} = -\partial L'/\partial \, _0 u_{i,j}$ is the effective nominal stress (In linear elasticity, we do not differentiate between the Cauchy stress and the second Piola-Kirchhoff stress.); ${}^\tau f_i^E = \partial L'/\partial u_i$ is the volume density of effective body force; and ${}^\tau p_i = \partial L'/\partial \dot{u}_i$ is the volume density of effective momentum.

While temporal inhomogeneity is intriguing and significant for certain applications [45-47, 67], it will take time for us to thoroughly grasp this concept. As such, we intend to set aside Eqs. (29cd) and (34) for the time being. Consequently, from now on we simply use the overhead dot to denote the time derivative, as exemplified by ${}^\tau \dot{p}_i$ in Eq. (36).

In accordance with the minimal replacement method, $L' = T' - W' - \Phi'$. Thus, by inserting Eq. (32) into Eqs. (7b) and (8), we obtain

$$\begin{aligned}T' &= {}^0\rho_{ij} \, {}^0\dot{x}_i \, _0 u_{j;t} + \frac{1}{2} {}^0\rho_{ij} \, _0 u_{i;t} \, _0 u_{j;t} \\ &= {}^0\rho_{ij}\left[{}^0\dot{x}_i\dot{u}_j + {}^\tau\dot{x}_i\left(_0\bar{\dot{u}}_{j,k} u_k - \bar{\dot{u}}_k \, _0 u_{j,k}\right) - {}_0\bar{\dot{u}}_{i,r}\bar{\dot{u}}_s \, _0 u_{j,s} u_r\right], \\ &\quad + \frac{1}{2}{}^0\rho_{ij}\left(\dot{u}_i\dot{u}_j + {}_0\bar{\dot{u}}_{i,r}\,_0\bar{\dot{u}}_{j,s} u_r u_s + \bar{\dot{u}}_r \bar{\dot{u}}_s \, _0 u_{i,r} \, _0 u_{j,s}\right)\end{aligned} \qquad (37)$$

$$\begin{aligned}W' &= {}^0\sigma_{ij} \, _0 u_{i;j} + \frac{1}{2} C_{ijkl} \, _0 u_{i;j} \, _0 u_{k;l} \\ &= {}^0\sigma_{ij}\left(_0 u_{i,j} + {}_0\bar{u}_{i,jk} u_k\right) + \frac{1}{2} C_{ijkl}\left(_0 u_{i,j} + {}_0\bar{u}_{i,jr} u_r\right)\left(_0 u_{k,l} + {}_0\bar{u}_{k,ls} u_s\right)\end{aligned}, \qquad (38)$$

$$\Phi' = \Phi = -\left({}^0f_i + \frac{1}{2}{}^0f_{i,j}u_j + \dot{f}_i\right)u_i. \tag{39}$$

## 3.2 Minimal Coupling Method

Eqs. (37) and (38) indicate that some energy is stored in the gauge field. Since we focus on infinitesimal perturbations from the equilibrium state in linear elasticity, we can employ the PMD [55] to calculate the energy contained in the gauge field. This approach aligns with the methodology of minimal coupling [6, 7, 53] and can be interpreted as a gauge condition for the determination of the gauge field.

During the evolution of a continuum, the gauge field for the current state is assembled from the velocities and strains of the preceding configuration. This is why the classical Willis equations, as presented in Eq. (1), incorporate temporal-spatial convolutions. Given that the strain energy has already been minimized in the earlier configuration, we can directly link the spatial gauge ${}_0\bar{u}_{i,j}$ to the pre-strain ${}^0e_{ij}$. I.e., ${}_0\bar{u}_{i,j} = {}^0_0u_{i,j}$, where ${}^0u_i$ denotes the pre-displacement and ${}^0_0u_{i,j}$ is defined through

$$ {}^0e_{ij} = \frac{1}{2}\left({}^0_0u_{i,j} + {}^0_0u_{j,i}\right). \tag{40}$$

Therefore, according to the general Hooke's law and the symmetry of $C_{ijkl}$, we have

$$ {}^0\sigma_{ij} = C_{ijkl}\,{}^0e_{ij} = C_{ijkl}\,{}_0\bar{u}_{k,l} = \frac{\partial\,{}^0W}{\partial\,{}_0\bar{u}_{i,j}}, \quad C_{ijkl} = \frac{\partial^2\,{}^0W}{\partial\,{}_0\bar{u}_{i,j}\partial\,{}_0\bar{u}_{k,l}}, \tag{41}$$

where ${}^0W$ is the volume density of strain energy in configuration ${}^0B$. Substituting Eq. (41) into Eq. (38), we obtain

$$W' = {}^0W_{,i}u_i + {}^0\sigma_{ij}\,{}_0u_{i,j} + \frac{1}{2}{}^0W_{,ij}u_iu_j + {}^0\sigma_{ij,k}\,{}_0u_{i,j}u_k + \frac{1}{2}C_{ijkl}\,{}_0u_{i,j}\,{}_0u_{k,l}. \tag{42}$$

After fixing the spatial gauge as ${}_0\bar{u}_{i,j} = {}^0_0u_{i,j}$, the temporal gauge ${}_0\bar{u}_{i,t}$ can be determined by minimizing the incompatibility tensor defined in Eq. (27). Substituting Eqs. (23), (24) and (30) into Eq. (27), and noticing ${}_0\bar{u}_{j,k} \ll 1$, we can obtain

$$\begin{aligned}r_{i\beta\mu} &= {}_0\bar{u}_{i,\mu\gamma}\,{}_0\left({}_0\bar{u}_{\gamma,\nu}\,{}^0u_\nu\right)_{,\beta} - {}_0\bar{u}_{i,\beta\gamma}\,{}_0\left({}_0\bar{u}_{\gamma,\nu}\,{}^0u_\nu\right)_{,\mu} + \left({}_0\bar{u}_{\beta,\mu\gamma} - {}_0\bar{u}_{\mu,\beta\gamma}\right){}_0\bar{u}_{i;\gamma} \\ &\quad + {}_0\bar{u}_{i,\gamma\mu}\,{}_0\bar{u}_{\gamma,\beta} - {}_0\bar{u}_{i,\gamma\beta}\,{}_0\bar{u}_{\gamma,\mu} + {}_0\bar{u}_{\beta,\gamma}\,{}_0\left({}_0\bar{u}_{i;\gamma}\right)_{,\mu} - {}_0\bar{u}_{\mu,\gamma}\,{}_0\left({}_0\bar{u}_{i;\gamma}\right)_{,\beta} \\ &\approx {}_0\bar{u}_{i,j\mu}\,{}_0\bar{u}_{j,\beta} - {}_0\bar{u}_{i,j\beta}\,{}_0\bar{u}_{j,\mu} + {}_0\left({}_0\bar{u}_{\beta,\gamma}\,{}_0\bar{u}_{i;\gamma}\right)_{,\mu} - {}_0\left({}_0\bar{u}_{\mu,\gamma}\,{}_0\bar{u}_{i;\gamma}\right)_{,\beta}\end{aligned} \tag{43}$$

Ignoring $\bar{u}_t$, the components in $r_{i\beta\mu}$ related to the temporal gauge $\dot{\bar{u}}_l$ are

$$r_{ilt} = -r_{itl} \approx F_{il} + \dot{H}_l\dot{\bar{u}}_l + H_l\ddot{\bar{u}}_l, \tag{44}$$

where $F_{il} = {}_0^0\dot{u}_{i,j}\,{}_0^0 u_{j,l} - {}_0^0 u_{i,jl}\,{}_0^0\dot{u}_j + \left({}_0^0 u_{l,j}\,{}_0^0 u_{i;j}\right)_{,t}$ and $H_i = {}_0^0 u_{i;t}$ are already known.

According to the PMD, the temporal gauge $\dot{\bar{u}}_l$ should minimize the incompatibility factor $R = r_{ilt} r_{ilt}$, i.e.,

$$\frac{\partial R}{\partial \dot{\bar{u}}_l} = 0 \quad \Rightarrow \quad \dddot{\bar{u}}_l + \frac{1}{\Lambda}\ddot{\bar{u}}_l + F_l = 0, \tag{45}$$

where $F_l = \dot{H}_i F_{il}/(H_i \dot{H}_i)$, and $\Lambda = H_i \dot{H}_i/(\dot{H}_i \dot{H}_i)$ is the characteristic time. The solution of Eq. (45) is

$$\dot{\bar{u}}_l = e^{-\int \frac{1}{\Lambda}\mathrm{d}t}\left(-\int F_l e^{\int \frac{1}{\Lambda}\mathrm{d}t}\,\mathrm{d}t + {}^0\dot{\bar{u}}_l\right). \tag{46}$$

This indicates that the evolution of $\dot{\bar{u}}_l$ is not time reversable, the symptom of dissipation. We will revisit this intriguing topic in Section 4.3.

## 3.3 Generalized Field Equations

Equipped with the established gauge fields and the Lagrangian $L'$ that incorporates Eqs. (37), (39), and (42), we are positioned to derive the following field equations:

$$^\tau\sigma_{ij} = -\frac{\partial L'}{\partial\,{}_0 u_{i,j}} = {}^0\sigma_{ij} + {}_0^0\sigma_{ij,k} u_k + C_{ijkl}\,{}_0 u_{k,l} + {}^0\rho_{ik}\dot{\bar{u}}_j\left({}_0\dot{\bar{u}}_{k,l} u_l - \dot{\bar{u}}_l\,{}_0 u_{k,l} + {}^\tau\dot{x}_k\right), \tag{47}$$

$$^\tau p_i = \frac{\partial L'}{\partial \dot{u}_i} = {}^0\rho_{ij}\left({}^\tau\dot{x}_j + {}_0\dot{\bar{u}}_{j,k} u_k - \dot{\bar{u}}_k\,{}_0 u_{j,k}\right), \tag{48}$$

$$^\tau f_i^E = \frac{\partial L'}{\partial u_i} = {}^0 f_i + {}_0^0 f_{i,j} u_j + f_i + {}^0\rho_{kj}\left({}^\tau\dot{x}_k + {}_0\dot{\bar{u}}_{k,l} u_l - \dot{\bar{u}}_l\,{}_0 u_{k,l}\right){}_0\dot{\bar{u}}_{j,i}$$
$$- {}_0^0\sigma_{kj,i}\,{}_0 u_{k,j} - {}_0^0 W_{,i} - {}_0^0 W_{,ij} u_j \,. \tag{49}$$

With Eq. (49), we know the effective volume density of body force in ${}^0 B$ is

$$^0 f_i^E = {}^0 f_i + {}^0\rho_{kj}\,{}^0\dot{x}_k\,{}_0\dot{\bar{u}}_{j,i} - {}_0^0 W_{,i}, \tag{50}$$

which is in balance with the pre-stress and inertial force in ${}^0 B$ (neglecting ${}^0\dot{\rho}_{ij}$)

$$^0_0\sigma_{ij,j} + {}^0 f_i^E = {}^0\dot{p}_i \approx {}^0\rho_{ij}\,{}^0\ddot{x}_j. \tag{51}$$

Substituting Eqs. (47)-(51) into Eq. (36), and neglecting ${}^0\dot{\rho}_{ij}$, ${}_0\dot{\bar{u}}_{l,ij}$, ${}_0\ddot{\bar{u}}_{j,k}$, ${}_0\ddot{x}_{k,j}$, we can obtain the equation of motion:

$$_0\sigma_{ij,j} + f_i \approx {}_0^0\sigma_{ik,kj} u_j + {}_0^0\sigma_{kj,i}\,{}_0 u_{k,j} + {}^0\rho_{kl}\,{}_0\dot{\bar{u}}_{l,i}\left[\left({}_0^0\dot{x}_{k,j} - {}_0\dot{\bar{u}}_{k,j}\right) u_j + \dot{\bar{u}}_j\,{}_0 u_{k,j}\right]$$
$$+ \left({}^0\rho_{ij}\,{}_0\dot{\bar{u}}_{j,k} - {}^0\rho_{kj}\,{}_0\dot{\bar{u}}_{j,i}\right)\dot{u}_k - {}^0\rho_{ij}\ddot{\bar{u}}_k\,{}_0\dot{u}_{j,k} + {}^0\rho_{ij}\ddot{u}_j \,, \tag{52}$$

where the incremental nominal stress is

$$\sigma_{ij} = {}^{\tau}\sigma_{ij} - {}^{0}\sigma_{ij} = \left({}^{0}_{0}\sigma_{ij,k} + {}^{0}\rho_{il}\dot{\bar{u}}_{j}{}_{0}\dot{\bar{u}}_{l,k}\right)u_{k} + \left(C_{ijkl} - {}^{0}\rho_{ik}\dot{\bar{u}}_{j}\dot{\bar{u}}_{l}\right){}_{0}u_{k,l} + {}^{0}\rho_{ik}\dot{\bar{u}}_{j}{}^{\tau}\dot{x}_{k}. \quad (53)$$

Beyond the classical Willis equations given in Eq. (1) and the Willis-form equations given in Eq. (2), Eqs. (52) and (53) include additional terms attributable to temporal-spatial coupling. It is noteworthy that ${}^{0}\rho_{ik}\dot{\bar{u}}_{j}$ in Eq. (53) and $-{}^{0}\rho_{ij}\dot{\bar{u}}_{k}$ in Eq. (52) do not have the symmetries of their counterparts $S^{eff}_{ijk} = S^{eff}_{jik}$ in Eq. (1a) and $S^{\dagger eff}_{ijk} = S^{\dagger eff}_{ikj}$ in Eq. (1b). This is because $S^{eff}_{ijk}$ is intentionally symmetrized, as depicted in Eq. (3.8) to Eq. (3.10) from reference [17]. However, as Eq. (4.3) from the original paper [16] indicates, it is not necessary to require $S^{eff}_{ijk} = S^{eff}_{jik}$ nor $S^{\dagger eff}_{ijk} = S^{\dagger eff}_{ikj}$.

Owing to the lack of symmetry in the coupling terms, the incremental stress $\sigma_{ij}$ in Eq. (53) is not symmetric, indicating the presence of localized rotational backgrounds. This aligns with the result from Eq. (52), which reveals a localized gyroscopic force $\left({}^{0}\rho_{ij}{}_{0}\dot{\bar{u}}_{j,k} - {}^{0}\rho_{kj}{}_{0}\dot{\bar{u}}_{j,i}\right)\dot{u}_{k}$ arising from the localized Coriolis effect due to the temporal gauge field $\dot{\bar{u}}_{i}$. Considering that relative velocity serves as the cause of viscosity, the presence of localized rotational backgrounds supports the established understanding that rotational motions are inherently linked to viscous effects.

## 4 Thermodynamics

### 4.1 Thermal properties

Thermodynamics is closely related to the concept of temperature, which represents the intensity of micro-vibrations. For example, the temperature $\theta$ of an ideal gas in equilibrium is correlated to the average kinetic energy $\overline{m\dot{u}_{i}\dot{u}_{i}}/2$ of gas molecules:

$$\frac{3}{2}k_{B}\theta = \frac{1}{2}\overline{m\dot{u}_{i}\dot{u}_{i}}, \quad (54)$$

where $k_{B}$ is the Boltzmann constant. For elastic media, micro-vibrations can be represented by the relative velocities among material points, i.e., the temporal gauge field $\dot{\bar{u}}_{j}$. Therefore, similar to Eq. (54) we can define the incremental temperature $\Delta\theta$ of a material point through

$$\frac{3}{2}k_{B}\Delta\theta\delta_{ij} = \frac{1}{2}{}^{0}\rho_{ik}\dot{\bar{u}}_{j}\dot{\bar{u}}_{l}{}_{0}u_{k,l}\bar{V}, \quad (55)$$

where $\bar{V}$ is the representative spatial domain of this material point. Eq. (55) is related to ${}^{0}\rho_{ik}\dot{\bar{u}}_{j}\dot{\bar{u}}_{l}{}_{0}u_{k,l}$ in Eq. (53). This term can be regarded as the stress due to anisotropic thermal expansions:

$$^{0}\rho_{ik}\dot{\bar{u}}_{j}\dot{\bar{u}}_{l\ 0}u_{k,l} = C_{ijkl}\alpha_{kl}\Delta\theta. \tag{56}$$

$\alpha_{kl}$ is thermal expansion tensor, which can be formulated using homogenization techniques as referenced in [68]. However, by merging Eqs. (55) and (56), we gain a straightforward understanding that:

$$\alpha_{kl} = \frac{3k_{B}C_{iikl}^{-1}}{\bar{V}}, \tag{57}$$

where $C_{iikl}^{-1}$ denotes the components of compliance tensor. Eq. (57) aligns with the established concept of the thermal expansion coefficient in solid-state physics [51].

Substituting Eq. (56) into Eq. (53), it yields

$$\sigma_{ij} = \left(_{0}^{0}\sigma_{ij,k} + {}^{0}\rho_{il}\dot{\bar{u}}_{j\ 0}\dot{\bar{u}}_{l,k}\right)u_{k} + C_{ijkl}\left(_{0}u_{k,l} - \alpha_{kl}\Delta\theta\right) + {}^{0}\rho_{ik}\dot{\bar{u}}_{j}\ {}^{\tau}\dot{x}_{k}. \tag{58}$$

In addition to the classical general Hooke's law, the extra terms in Eq. (58) account for the contributions from the pre-stress gradient and viscous effects including ${}^{0}\rho_{il}\dot{\bar{u}}_{j\ 0}\dot{\bar{u}}_{l,k}u_{k}$, ${}^{0}\rho_{ik}\dot{\bar{u}}_{j}$ and the thermal strain $\alpha_{kl}\Delta\theta$.

## 4.2 The First Law of Thermodynamics

Multiplying Eq. (52) with velocity $\dot{u}_{i}$, and integrating over a spatial domain $V$, we can obtain

$$\int_{V}\left(_{0}\sigma_{ij,j} + f_{i}\right)\dot{u}_{i}\mathrm{d}V$$
$$\approx \int_{V}\left(_{0}^{0}\sigma_{ik,kj}u_{j} + {}_{0}^{0}\sigma_{kj,i\ 0}u_{k,j}\right)\dot{u}_{i}\mathrm{d}V + \int_{V}{}^{0}\rho_{kl\ 0}\dot{\bar{u}}_{l,i}\left[\left(_{0}^{0}\dot{x}_{k,j} - {}_{0}\dot{\bar{u}}_{k,j}\right)u_{j} + \dot{\bar{u}}_{j\ 0}u_{k,j}\right]\dot{u}_{i}\mathrm{d}V \ .$$
$$+ \int_{V}\left(^{0}\rho_{ij\ 0}\dot{\bar{u}}_{j,k} - {}^{0}\rho_{kj\ 0}\dot{\bar{u}}_{j,i}\right)\dot{u}_{k}\dot{u}_{i}\mathrm{d}V - \int_{V}{}^{0}\rho_{ij}\dot{\bar{u}}_{k\ 0}\dot{u}_{j,k}\dot{u}_{i}\mathrm{d}V + \int_{V}{}^{0}\rho_{ij}\ddot{u}_{j}\dot{u}_{i}\mathrm{d}V$$

Applying integration by parts, the above equation becomes

$$\int_{V}{}^{0}\rho_{ij}\ddot{u}_{j}\dot{u}_{i}\mathrm{d}V + \int_{V}{}^{0}\rho_{kl\ 0}\dot{\bar{u}}_{l,i}\left[\left(_{0}^{0}\dot{x}_{k,j} - {}_{0}\dot{\bar{u}}_{k,j}\right)u_{j} + \dot{\bar{u}}_{j\ 0}u_{k,j}\right]\dot{u}_{i}\mathrm{d}V$$
$$-\int_{V}{}^{0}\rho_{ij}\dot{\bar{u}}_{k\ 0}\dot{u}_{j,k}\dot{u}_{i}\mathrm{d}V + \int_{V}\left(_{0}^{0}\sigma_{ik,kj}u_{j} + {}_{0}^{0}\sigma_{kj,i\ 0}u_{k,j}\right)\dot{u}_{i}\mathrm{d}V + \int_{V}\sigma_{ij\ 0}\dot{u}_{i,j}\mathrm{d}V, \tag{59}$$
$$\approx \int_{V}f_{i}\dot{u}_{i}\mathrm{d}V + \int_{\partial V}\sigma_{ij}n_{i}\dot{u}_{i}\mathrm{d}S$$

where $\partial V$ denotes the boundary of $V$; and $n_{i}$ is the outward normal vector of $\partial V$.

Eq. (59) depicts *the principle of energy conservation*. The right-hand side represents the power exerted by external forces, including the incremental body force $f_{i}$ and the boundary traction $\sigma_{ij}n_{i}$. The left-hand side represents the power of internal forces, encompassing contributions from the inertial force ${}^{0}\rho_{ij}\ddot{u}_{j}$, the thermal expansion force ${}^{0}\rho_{kl\ 0}\dot{\bar{u}}_{l,i}\left[\left(_{0}^{0}\dot{x}_{k,j} - {}_{0}\dot{\bar{u}}_{k,j}\right)u_{j} + \dot{\bar{u}}_{j\ 0}u_{k,j}\right]$, the viscous damping force $-{}^{0}\rho_{ij}\dot{\bar{u}}_{k\ 0}\dot{u}_{j,k}$, the residual force ${}_{0}^{0}\sigma_{ik,kj}u_{j} + {}_{0}^{0}\sigma_{kj,i\ 0}u_{k,j}$, and the force due to incremental stress $\sigma_{ij}$. The appearance of unconventional forces in Eq. (59) is not unexpected, as it has been demonstrated that nonstandard terms can emerge in the transformed balance of energy even within the context of a Riemannian manifold [69].

Given that the power exerted by the gyroscopic force $\left( {}^0\rho_{ij}\, {}_0\dot{\bar{u}}_{j,k} - {}^0\rho_{kj}\, {}_0\dot{\bar{u}}_{j,i} \right)\dot{u}_k$ is null, we cannot find its contribution in Eq. (59). Nonetheless, the manifestation of localized rotational backgrounds may provide valuable insights into the design of topological metamaterials featured with spontaneous non-reciprocity [31].

The preceding analysis confirms that the gauge theory complies with the first law of thermodynamics. However, a fraction of energy is incorporated into the gauge field. This improper energy conservation law can be rationalized by applying the Noether's second theorem [70], as elaborated in Appendix B.

## 4.3 The Second Law of Thermodynamics

Eq. (58) manifests that the pre-stress, the thermal stress and the viscous stress are naturally included in this generalized theory of linear elasticity. The pre-stress is the result of spatial gauge ${}_0\bar{u}_{i,j}$, and is likely induced by micro-defects such as dislocations and damage, which are the main focus of geometric continuum mechanics [6-13]. The thermal stress and the viscous stress are the results of temporal gauge ${}_0\bar{u}_{i,t}$. All these stresses are originated from $\bar{u}_i$, with its temporal-spatial gradients characterizing the local interactions within the inhomogeneous medium. These interactions are the origin of dissipative effect, rendering them irreversible, as evidenced in Eqs. (45) and (46). Consequently, the forces associated with these interactions, as detailed in Eqs. (49) and (59), are certainly nonconservative.

The concepts of irreversibility and nonconservatism are distinct. In his influential work [54], Lanczos elucidated that the PLA is applicable to systems where forces can be derived from a scalar work function. He termed this kind of force as "monogenic". Monogenic forces are conservative when the work function is not time-dependent; otherwise, they are nonconservative. Lanczos also identified forces as "polygenic" if they cannot be derived from a work function, such as friction. He highlighted that polygenic forces "*do not yield to the general minimizing procedures of analytical mechanics.*"

However, Feynman clearly stated in the classical book [80]: "*As a matter of fact, all the fundamental forces in nature appear to be conservative. This is not a consequence of Newton's laws. In fact, so far as Newton himself knew, the forces could be nonconservative, as friction apparently is. When we say friction apparently is, we are taking a modern view, in which it has been discovered that all the deep forces, the forces between the particles at the most fundamental level, are* *conservative.*" Based on this viewpoint, it becomes feasible to go beyond the theoretical boundaries established in Lanczos' book, thereby enabling the application of the PLA to broader physical systems.

As shown in earlier discussions, the PLA can be extended to dissipative systems by employing the minimal replacement method to include gauge fields within the Lagrangian. This approach can be regarded as the "modern view" of Feynman's statement.

If all forces are conservative, the origin of irreversibility becomes a fundamental question that can be traced back to the critiques of the Boltzmann equation, as elaborated in reference [81]. In Boltzmann's model for an ideal gas, the molecules are considered to be hard, smooth spheres, implying that there is no energy loss during molecular collisions. However, the number of possible interaction sequences among these molecules is extraordinarily large. Given that only one of these sequences is reversible, the macroscopic evolution of the system is likely to be irreversible. This reasoning is analogous to the inhomogeneous media discussed in this paper. As depicted in Fig. 1, the interaction between two associated material points *X* and *X'* does not reduce the total energy but redistributes it due to the change of $\bar{u}_\alpha$. The sequence of this process is likely irreversible, manifesting a dissipative effect that is intimately connected to the second law of thermodynamics.

## 5 Concluding Remarks

In the concluding remarks of our previous paper [1], we envisaged the possibility of deriving new equations by considering temporal and spatial transformations simultaneously. This assertion has been realized and is elaborated upon in this paper. It is particularly intriguing to observe that various phenomena, including thermal stresses, damping, and local gyroscopic forces, are inherently incorporated within these newly formulated equations. The harmonious integration of these concepts underscores the efficacy of gauge theories.

This paper demonstrates that the PLA can be extended to dissipative systems upon identifying an appropriate Lagrangian, as illustrated by Eqs. (37) to (39). It is crucial to note that while the geometric aspect of this methodology is well-established, the material aspect is somewhat phenomenological. For instance, the strain energy in Eq. (38) is based on the general Hooke's law of simple materials. To broaden the scope of this theory to encompass other materials, such as materials with evolving damage [14, 15], innovative formulations are imperative.

Furthermore, this paper focuses on linear elasticity, so that it only accounts for small perturbations. For problems involving finite deformations, Eq. (32) could not be accurate enough, necessitating the inclusion of additional terms in Eq. (31).

*The principle of generalized objectivity* [7], which examines the local symmetries within a system, serves as a powerful tool for investigating the complexities of interactions. The gauge theories associated with this principle, including the PLA, minimal replacement, and minimal coupling techniques, have achieved notable successes in particle physics [53, 60, 61]. The gauge theory for linear elasticity introduced in this paper is just an exploratory step in continuum mechanics. It is anticipated that the gauge theory could play a more significant role in deepening our comprehension of this traditional field. For example, a recent study has revealed that the Navier-Stokes equations lack form-invariance due to their viscous terms [82], a finding that is closely related to the investigative methods employed in the analysis of the classical Willis equations [25].

**Acknowledgements** Professor John Willis opened the door and showed me a new world. Professor Gexue Ren reminded me to pay more attention to dissipation. Dr. Jing Xiao helped me in clarifying many fundamental concepts in theoretical physics. I am also grateful to many friends for their unwavering encouragement, which is instrumental in completing this work.

## Appendix A: Coordinate Transformation of $G_{\alpha\beta}$

Similar to Eq. (16), the covariant derivative of the transformed displacement is

$$_\tau u'_{\alpha;\beta'} = {_\tau u'_{\alpha,\beta'}} + G'_{\alpha\beta}. \tag{A1}$$

Substituting Eqs. (15) and (16) into Eq. (A1), it yields

$$\begin{aligned} {_\tau^0 x_{\gamma,\alpha'}} \left( {_0 u_{\gamma,\mu}} + G_{\gamma\mu} \right) {_\tau^0 x_{\mu,\beta'}} &= {_0\left( {_\tau^0 x_{\gamma,\alpha'}} u_\gamma \right)_{,\mu}} {_\tau^0 x_{\mu,\beta'}} + G'_{\alpha\beta} \\ G'_{\alpha\beta} &= \left[ {_\tau^0 x_{\gamma,\alpha'}} G_{\gamma\mu} - {_0\left( {_\tau^0 x_{\gamma,\alpha'}} \right)_{,\mu}} u_\gamma \right] {_\tau^0 x_{\mu,\beta'}} \end{aligned}. \tag{A2}$$

Under pure translation, $G'_{\alpha\beta} = \Gamma'^{\eta}_{\alpha\beta} u'_\eta$ and $G_{\gamma\mu} = \Gamma^{\nu}_{\gamma\mu} u_\nu$ according to Eq. (23). Substituting these terms and Eq. (13) into Eq. (A2), we can obtain

$$\Gamma'^{\eta}_{\alpha\beta} u'_\eta = \left( \Gamma^{\nu}_{\gamma\mu} {_\tau^0 x_{\gamma,\alpha'}} {_\tau^0 x_{\mu,\beta'}} {_0^\tau x_{\eta,\nu}} - {_\tau^0 x_{\gamma,\alpha'\beta'}} {_0^\tau x_{\eta,\gamma}} \right) u'_\eta. \tag{A3}$$

Since $u'_\eta$ is an arbitrary variable, Eq. (A3) becomes the coordinate transformation law for the classical connection

$$\Gamma'^{\eta}_{\alpha\beta} = \Gamma^{\nu}_{\gamma\mu} {_\tau^0 x_{\gamma,\alpha'}} {_\tau^0 x_{\mu,\beta'}} {_0^\tau x_{\eta,\nu}} - {_\tau^0 x_{\gamma,\alpha'\beta'}} {_0^\tau x_{\eta,\gamma}}. \tag{A4}$$

## Appendix B: Conservation Laws

### B.1 General Formula

Noether's theorems establish a relationship between conservation laws and the invariance or symmetry of the Hamiltonian action under infinitesimal transformations [70]. As discussed in Section 2.1, the variations of the coordinate and the displacement under these transformations can be specified as follows:

$$\delta x_\alpha = {}^\tau x'_\alpha - {}^0 x_\alpha, \tag{B1}$$

$$\delta u_\alpha = {}_0 u_{\alpha;\beta} \delta {}^0 x_\beta = \left({}_0 u_{\alpha,\beta} + G_{\alpha\beta}\right)\delta {}^0 x_\beta. \tag{B2}$$

Thus, we have

$$d\Omega' = \det\left({}^\tau_0 x'_{\alpha,\beta}\right) d\Omega \approx \left[1 + {}_0\left(\delta x_\gamma\right)_{,\gamma}\right] d\Omega, \tag{B3}$$

$$_\tau u'_{\alpha,\beta'} = {}_0 u'_{\alpha,\gamma} {}^0_\tau x_{\gamma,\beta'} \approx {}_0\left(u_\alpha + \delta u_\alpha\right)_{,\gamma}\left[\delta_{\gamma\beta} - {}_0\left(\delta x_\gamma\right)_{,\beta}\right]. \tag{B4}$$

Substituting Eqs. (B3) and (B4) into Eqs. (9) and (14), it yields

$$\int_\Omega L'\left({}^0 x_\alpha, u_\alpha, {}_0 u_{\alpha,\beta}\right) d\Omega$$

$$\approx \int_\Omega L'\left({}^0 x_\alpha + \delta x_\alpha, u_\alpha + \delta u_\alpha, {}_0 u_{\alpha,\beta} + {}_0\left(\delta u_\alpha\right)_{,\beta} - {}_0 u_{\alpha,\gamma} {}_0\left(\delta x_\gamma\right)_{,\beta}\right)\left[1 + {}_0\left(\delta x_\gamma\right)_{,\gamma}\right] d\Omega$$

$$\int_\Omega \left\{ {}_0 L'_{,\alpha}\delta x_\alpha + \frac{\partial L'}{\partial u_\alpha}\delta u_\alpha + \frac{\partial L'}{\partial {}_0 u_{\alpha,\beta}}\left[ {}_0\left(\delta u_\alpha\right)_{,\beta} - {}_0 u_{\alpha,\gamma} {}_0\left(\delta x_\gamma\right)_{,\beta}\right] + L' {}_0\left(\delta x_\gamma\right)_{,\gamma}\right\} d\Omega \approx 0. \tag{B5}$$

Referring to Eq. (33), Eq. (B5) can be rewritten as

$$\int_\Omega \left[ {}_0 L'_{,\alpha}\delta x_\alpha + E_\alpha \delta u_\alpha + \left(\frac{\partial L'}{\partial {}_0 u_{\alpha,\beta}}\delta u_\alpha\right)_{,\beta} - \frac{\partial L'}{\partial {}_0 u_{\alpha,\beta}} {}_0 u_{\alpha,\gamma} {}_0\left(\delta x_\gamma\right)_{,\beta} + L' {}_0\left(\delta x_\gamma\right)_{,\gamma}\right] d\Omega \approx 0, \tag{B6}$$

where $E_\alpha$ is not necessarily zero, and

$$-\frac{\partial L'}{\partial {}_0 u_{\alpha,\beta}} {}_0 u_{\alpha,\gamma} {}_0\left(\delta x_\gamma\right)_{,\beta}$$

$$= -\left(\frac{\partial L'}{\partial {}_0 u_{\alpha,\beta}} {}_0 u_{\alpha,\gamma}\delta x_\gamma\right)_{,\beta} + \left(\frac{\partial L'}{\partial {}_0 u_{\alpha,\beta}}\right)_{,\beta} {}_0 u_{\alpha,\gamma}\delta x_\gamma + \frac{\partial L'}{\partial {}_0 u_{\alpha,\beta}} {}_0 u_{\alpha,\beta\gamma}\delta x_\gamma$$

$$= -\left(\frac{\partial L'}{\partial {}_0 u_{\alpha,\beta}} {}_0 u_{\alpha,\gamma}\delta x_\gamma\right)_{,\beta} + \left[\left(\frac{\partial L'}{\partial u_\alpha} - E_\alpha\right) {}_0 u_{\alpha,\gamma} + \frac{\partial L'}{\partial {}_0 u_{\alpha,\beta}} {}_0 u_{\alpha,\beta\gamma}\right]\delta x_\gamma \tag{B7}$$

$$\approx -\left(\frac{\partial L'}{\partial {}_0 u_{\alpha,\beta}} {}_0 u_{\alpha,\gamma}\delta x_\gamma\right)_{,\beta} + \left(\frac{dL'}{d{}^0 x_\gamma} - {}_0 L'_{,\gamma} - E_\alpha {}_0 u_{\alpha,\gamma}\right)\delta x_\gamma$$

Substituting Eqs. (B2) and (B7) into Eq. (B6), we can obtain

$$\int_\Omega {}_0 J_{\beta,\beta} d\Omega = -\int_\Omega E_\alpha\left(\delta u_\alpha - {}_0 u_{\alpha,\beta}\delta x_\beta\right) d\Omega = -\int_\Omega E_\alpha G_{\alpha\beta}\delta x_\beta d\Omega, \tag{B8}$$

where $J_\beta$ is the Noether current [71]:

$$J_\beta = \left(L'\delta_{\beta\gamma} + \frac{\partial L'}{\partial {}_0 u_{\alpha,\beta}} G_{\alpha\gamma}\right)\delta x_\gamma. \tag{B9}$$

## B.2 Proper Conservation Laws

For homogeneous media, the gauge field is either null or a constant, resulting in $G_{\alpha\gamma} = 0$ and $L' = L$. In addition, under a global transformation, $\delta x_\gamma$ is linearly dependent on $N$ infinitesimal constant parameters $\xi_i$ ($i = 1, 2, \ldots, N$). In this case, and taking $E_\alpha = 0$, Eqs. (B8) and (B9) simplify to

$$\int_{_0\Omega} \left[ L \frac{\partial(\delta x_\beta)}{\partial \xi_i} \right]_{,\beta} \mathrm{d}\Omega = 0. \tag{B10}$$

Eq. (B10) is the Noether's first theorem, which is associated with proper conservation laws. For example, under global translation, Eq. (B10) becomes $_0L_{,\beta} = 0$, implying that $L$ is not explicitly dependent on coordinate $^0x_\beta$. Therefore, referring to Eq. (33), we have

$$\frac{\mathrm{d}L}{\mathrm{d}\,^0x_\beta} = \frac{\partial L}{\partial u_\alpha}\,_0u_{\alpha,\beta} + \frac{\partial L}{\partial\,_0u_{\alpha,\gamma}}\,_0u_{\alpha,\gamma\beta} = \frac{\mathrm{d}}{\mathrm{d}\,^0x_\gamma}\left( \frac{\partial L}{\partial\,_0u_{\alpha,\gamma}}\,_0u_{\alpha,\beta} \right). \tag{B11}$$

This means

$$\int_\Omega {_0T_{\beta\gamma,\gamma}}\,\mathrm{d}\Omega = 0, \quad T_{\beta\gamma} = L\delta_{\beta\gamma} - \frac{\partial L}{\partial\,_0u_{\alpha,\gamma}}\,_0u_{\alpha,\beta}. \tag{B12}$$

$T_{\beta\gamma}$ is the generalized energy-momentum tensor, which degenerates to the Eshelby energy-momentum tensor (or Eshelby stress) [72], when disregarding time inhomogeneity and neglecting dynamical terms. The dynamical version of Eshelby energy-momentum tensor is also discussed in reference [73].

Ignoring temporal inhomogeneity and focusing solely on temporal translation (set $\beta = t$), Eq. (B12) suggests

$$\frac{\mathrm{d}}{\mathrm{d}t}\left[ \int_V \left( {^\tau p_i \dot{u}_i} - L \right) \mathrm{d}V \right] = \int_{\partial V} {^\tau \sigma_{ij} n_j \dot{u}_i}\,\mathrm{d}S. \tag{B13}$$

According to Eqs. (7b) and (8), we know ${^\tau p_i \dot{u}_i} - L = T + \Pi - {^0\rho_{ij}}\,{^0\dot{x}_i \dot{u}_j}$. Therefore, moving the contribution of inertial force ${^0\rho_{ij}}\,{^0\dot{x}_i \dot{u}_j}$ to the right-hand side of Eq. (B13), we can derive the condition of energy evolution.

Ignoring temporal inhomogeneity and focusing solely on spatial translation (set $\beta = k$), Eq. (B12) suggests

$$\frac{\mathrm{d}}{\mathrm{d}t}\left( \int_V {^\tau p_i\,_0u_{i,k}}\,\mathrm{d}V \right) = \int_{\partial V} \left( L n_k + {^\tau\sigma_{ij} n_j\,_0u_{i,k}} \right)\mathrm{d}S. \tag{B14}$$

The concept of pseudo-momentum ${^\tau p_i\,_0u_{i,k}}$ is thoroughly examined in reference [73]. Eq. (B14) was applied to address the issue of dynamic crack propagation as detailed in reference [74]. By disregarding the dynamical components, it simplifies to $\int_{\partial V}\left[(W+\Phi)n_k - {}^\tau\sigma_{ij}n_j{}_0u_{i,k}\right]\mathrm{d}S = 0$, leading to the path-independent *J*-integral.

Following the similar procedure, we can obtain more conservation laws related to a broader range of global transformations such as rotation and scaling, among others [75].

## B.3 Improper Conservation Laws

Inhomogeneous media are featured by the existence of gauge field, which describes local interactions among material points. In this case, the $G_{\alpha\beta}\delta x_\beta$ in Eq. (B8) varies at different locations, corresponding to local transformations. Thus, by consulting Eq. (23) and not insisting on $E_\alpha = 0$, we can obtain:

$$E_\alpha G_{\alpha\beta}\delta x_\beta = E_\alpha {}_0\left({}_0\bar{u}_{\alpha,\gamma}u_\gamma\right)_{,\beta}\delta x_\beta$$
$$= {}_0\left(E_\alpha {}_0\bar{u}_{\alpha,\gamma}u_\gamma\delta x_\beta\right)_{,\beta} + \left[E_\alpha {}_0 u_{\gamma,\beta}\delta x_\beta - {}_0\left(E_\alpha u_\gamma\delta x_\beta\right)_{,\beta}\right]{}_0\bar{u}_{\alpha,\gamma} \quad . \tag{B15}$$

Substituting Eq. (B15) into Eq. (B8), it yields

$$\int_{\Omega}{}_0\left(J_\beta + E_\alpha {}_0\bar{u}_{\alpha,\gamma}u_\gamma\delta x_\beta\right)_{,\beta}\mathrm{d}\Omega = \int_{\Omega}\left[{}_0\left(E_\alpha u_\gamma\delta x_\beta\right)_{,\beta} - E_\alpha {}_0 u_{\gamma,\beta}\delta x_\beta\right]{}_0\bar{u}_{\alpha,\gamma}\mathrm{d}\Omega. \tag{B16}$$

Given that the left-hand side of Eq. (B16) vanishes on the boundary, we arrive at the Noether's second theorem as follows [70, 71]:

$${}_0\left(E_\alpha u_\gamma\delta x_\beta\right)_{,\beta} = E_\alpha {}_0 u_{\gamma,\beta}\delta x_\beta. \tag{B17}$$

Since Eq. (B17) is independent of the equation of motion ($E_\alpha = 0$), the solutions to this problem seem not unique [71]. This non-uniqueness is related to the concept of gauge freedom. As elaborated in Section 3.2, the gauge field can be determined based on the PMD [55]. By implementing those gauge conditions, we can derive unique solutions that have distinct physical significance.

The vanishing of the left-hand side of Eq. (B16) implies the existence of a potential $\Theta_\beta$, whose divergence is always zero. I.e.,

$$\Theta_\beta = J_\beta + E_\alpha {}_0\bar{u}_{\alpha,\gamma}u_\gamma\delta x_\beta = \varepsilon_{\beta\gamma\alpha}q_{\gamma,\alpha}, \tag{B18}$$

where $q_\gamma$ is an arbitrary vector. Hence, the Noether current can be represented as:

$$J_\beta = \Theta_\beta - E_\alpha {}_0\bar{u}_{\alpha,\gamma}u_\gamma\delta x_\beta. \tag{B19}$$

Since ${}_0\Theta_{\beta,\beta} = 0$, we just obtain a trivial identity ${}_0 J_{\beta,\beta} = 0$ when $E_\alpha = 0$. If the global transformations discussed in Section B.2 are specific instances of local transformations, then this identity implies that classical conservation laws are inherently trivial for inhomogeneous media. These are called improper conservation laws by Hilbert, when he concerned about the lack of local energy conservation in general theory of relativity [76-78]. To address this matter, Noether published her seminal paper in 1918 [70], highlighting that the existence of improper conservation laws is the characteristic of a system with local symmetries. In such scenarios, we should set $L = L'$ in Eq. (B12), and obtain

$$\int_{\Omega_0} \left( T_{\beta\gamma} + t_{\beta\gamma} \right)_{,\gamma} \mathrm{d}\Omega = 0, \tag{B20}$$

where $t_{\beta\gamma}$ is associated with $G_{\alpha\beta}$. Because $G_{\alpha\beta}$ is not a real tensor as discussed in Appendix A, $t_{\beta\gamma}$ is called pseudo-tensor [76, 77]. Eq. (B20) reveals that the property of path independence, which is the characteristic of certain classical indices like *J*-integral, does not hold for inhomogeneous media [79] due to $t_{\beta\gamma}$. However, the analysis in Section 4.2 elucidates that energy conservation can be ensured for inhomogeneous media, albeit with some energy allocation to the gauge field. This forms the basis of our comprehension of improper conservation laws.